\documentclass[aps,pre,reprint,groupedaddress,showpacs,twocolumn]{revtex4}
\usepackage {graphicx}
\usepackage{amsmath}

\begin{document}

\title{Phase transition of the six-state clock model observed from the entanglement entropy}

\author{Roman Kr\v{c}m\'ar$^{1}$}
\author{Andrej Gendiar$^{1}$}
\email[]{andrej.gendiar@savba.sk}
\author{Tomotoshi Nishino$^{2}$}
\affiliation{$^1$Institute of Physics, Slovak Academy of Sciences, D\'ubravsk\'a cesta 9, SK-845 11, Bratislava, Slovakia}
\affiliation{$^2$Department of Physics, Graduate School of Science, Kobe University, Kobe 657-8501, Japan}

\date{\today}

\begin{abstract}
The Berezinskii-Kosterlitz-Thouless (BKT) transitions of the six-state clock model on the square lattice are 
investigated by means of the corner-transfer matrix renormalization group method. The classical analogue 
of the entanglement entropy $S( L, T )$ is calculated for $L$ by $L$ square system up to $L = 129$, as
a function of temperature $T$. The entropy has a peak at $T = T^{*}_{~}( L )$, where the temperature  
depends on both $L$ and boundary conditions. Applying the finite-size scaling to $T^{*}_{~}( L )$ and
assuming the presence of BKT transitions, the transition temperature is estimated to be $T_1^{~} = 0.70$ and
$T_2^{~} = 0.88$. The obtained results agree with previous analyses. It should be noted that no thermodynamic
function is used in this study.
\end{abstract}

\maketitle

\section{Introduction}

The classical XY model on uniform planar lattices does not exhibit a `standard type' of order when the 
temperature $T$ is finite, since the system possesses the continuous $O(2)$ symmetry~\cite{mermin}.
A special type of order that does not break the symmetry can, however, exist at finite temperature and is
known as the topological order~\cite{berezinskii, kosterlitz}. The phase transition between this topological 
phase and the high-temperature paramagnetic (or disordered) phase is the so-called
Berezinskii-Kosterlitz-Thouless (BKT) phase transition.

Introduction of anisotropy or discreteness is relevant to the thermodynamic properties of the system. 
The $q$-state clock model is one of the well-known examples, where on each lattice point there 
is a vector spin pointing to $q$ different directions, which differ by the angle $2\pi / q$. Since there is
no continuous symmetry, existence of standard ferromagnetic order is allowed at low but finite 
temperature. An early renormalization-group (RG) study on such a system by Jos\'e and Kadanoff suggested 
existence of a critical area with a finite temperature width~\cite{jose}, which is separated from ordered 
and disordered phases by the BKT phase transition~\cite{berezinskii,kosterlitz}. It has been known that 
such a phase structure exists for ferromagnetic $q$-state clock models when $q \ge 5$. 

In this article we consider the ferromagnetic six-state ($q = 6$) clock model on the square 
lattice, as a representative case where the BKT transition can be observed. The Hamiltonian 
of the system is written as
\begin{equation}
	H = - J \sum_{i,j=1}^L \left[
		\cos\left( \theta_{i,j}^{~} - \theta_{i+1,j}^{~} \right)
	      + \cos\left( \theta_{i,j}^{~} - \theta_{i,j+1}^{~} \right)
	                       \right] \, ,
\end{equation}
where $J > 0$ represents the bond strength between the nearest-neighboring spins on the square lattice
$L \times L$, and where $\theta_{i,j}^{~}=\frac{2\pi k}{q}$ denotes a discrete angle variable for
$k=0,1,2,\dots,q-1$ on the lattice site with coordinates $i$ and $j$. When the temperature $T / J$
is large enough, the thermal equilibrium state 
is disordered, and each direction is equally chosen. When $T / J$ is small enough, the state is ordered, 
where one of the six directions is spontaneously chosen in the thermodynamic limit. In the following we 
set $J = 1$ for convenience. 

It is known that within a temperature region $T_1^{~} < T < T_2^{~}$
the correlation function shows a power-law decay, and the system is critical. 
Table~I shows the value of $T_1^{~}$ and $T_2^{~}$, which have been reported so far
for the six-state clock model~\cite{tobochnik,challa,yamagata,tomita,hwang,brito,baek,kumano}.
In the early stage of the study, the finite-size corrections are treated according to simple
power-law scalings. Tobochnik performed Monte-Carlo simulation up to the system size $L = 32$. 
From behavior of the specific heat, the values $T_1^{~} = 0.6$ and $T_2^{~} = 1.3$ 
were estimated~\cite{tobochnik}.
Challa and Landau treated systems up to $L = 72$, and from the finite-size scaling on their 
Monte-Carlo result, they obtained $T_1^{~} = 0.68(2)$ and $T_2^{~} = 0.92(1)$.~\cite{challa}
Yamagata and Ono focused on the helicity modulus, and estimated $T_1^{~} = 0.68$ and 
$T_2^{~} = 0.90$ based on the data up to $L = 20$~\cite{yamagata}.

\begin{table}[b]
\begin{tabular}{lll}
\hline\hline
~ & $T_1^{~}$ & $T_2^{~}$ \\
\hline
Tobochnik~\cite{tobochnik}{\tiny (1982)} & 0.6 & 1.3 \\
Challa and Landau~\cite{challa}{\tiny (1986)} & 0.68(2) & 0.92(1) \\
Yamagata and Ono~\cite{yamagata}{\tiny (1991)} & 0.68 & 0.90 \\
\hline
Tomita and Okabe~\cite{tomita}{\tiny (2002)} & 0.7014(11) & 0.9008(6) \\
Hwang~\cite{hwang}{\tiny (2009)} & 0.632(2) & 0.997(2) \\
Brito {\it et al.\,}~\cite{brito}{\tiny (2010)} & 0.68(1) & 0.90(1) \\
Baek {\it et al.\,}~\cite{baek}{\tiny (2013)} & - & 0.9020(5) \\
Kumano {\it et al.\,}~\cite{kumano}{\tiny (2013)} & 0.700(4) & 0.904(5) \\
\hline
current work & 0.70 & 0.88  \\
\hline\hline
\end{tabular}
\caption{\footnotesize 
List of lower and upper BKT transition temperatures $T_1^{~}$ and
$T_2^{~}$, respectively, reported so far. Logarithmic scaling functions are
used after Tomita and Okabe~\cite{tomita}.}
\end{table}

An extensive calculation was performed by Tomita and Okabe for $L \le 512$ who reported
$T_1^{~} = 0.7014(11)$ and $T_2^{~} = 0.9008(6)$~\cite{tomita}.
From observation of Fisher zeros, Huwang estimated that $T_1^{~} = 0.632(2)$ and 
$T_2^{~} = 0.997(2)$ from relatively small systems with $L \le 28$, whereas fitting by BKT scaling 
form draws $T_1^{~} = 0.74$ and $T_2^{~} = 0.88$~\cite{hwang}.
Brito {\it et al.} treated the system as a class of random surface models and reported
$T_1^{~} = 0.68(1)$ and $T_2^{~} = 0.90(1)$ for $L \le 60$~\cite{brito}.
Baek {\it et al.} reported the value $T_2^{~} = 0.9020(5)$ from the helicity module~\cite{baek}.
One of the latest result is $T_1^{~} = 0.700(4)$ and $T_2^{~} = 0.904(5)$ by Kumano
{\it et al.}, based on the response to twist boundary conditions up to $L = 256$~\cite{kumano}.

Entanglement entropy, which quantifies the bipartite quantum entanglement, is one of the 
fundamental values in information physics, and has been used for analyses of one-dimensional (1D) 
quantum systems~\cite{osborne, vidal, franchini}.
Through the quantum-classical correspondence formulated by means of discrete path-integral in
imaginary time, such as the Trotter-Suzuki decomposition~\cite{trotter, suzuki}, it is also possible 
to introduce a classical analogue of the entanglement entropy for two-dimensional (2D) classical 
lattice systems~\cite{tagliacozzo}. A profit of using this classical analogue is that it enables to
detect thermal phase transitions directly, without considering the type of order parameter or without
taking derivatives of thermodynamic functions including the free energy~\cite{krcmar1, krcmar2, krcmar3}. 
Universality of the phase transition 
can also be identified by estimating the central charge through the finite-entanglement 
scaling~\cite{tagliacozzo,krcmar1, krcmar3}.

In this article we have observed the entanglement entropy around the BKT phase transition, in addition
to the first- and the second-order phase transitions considered in the context of entanglement entropy
so far. We focus on the scaling form of the entropy and on the numerical precision in the estimated
transition temperatures. In the following, we calculate the entanglement entropy $S( L, T )$ of the 
six-state cock model on square lattice systems of linear sizes $L$ up to $L = 129$ and investigate the
phase  transition by means of temperature dependence in $S( L, T )$. For this purpose we employ the 
corner-transfer-matrix renormalization group (CTMRG) method~\cite{tomotoshi}, which is based 
on Baxter's corner-transfer matrix (CTM) formalism~\cite{baxter}.

In the next section we briefly explain the construction of the density matrix by means of CTM, 
and introduce the classical analogue of the entanglement entropy $S( L, T )$. Numerical results 
are shown in Sec.~III. We summarize the obtained results and discuss relation between lattice 
geometry and nature of the phase transition.

\section{Entanglement Entropy}

\begin{figure}
\includegraphics[width=0.2\textwidth]{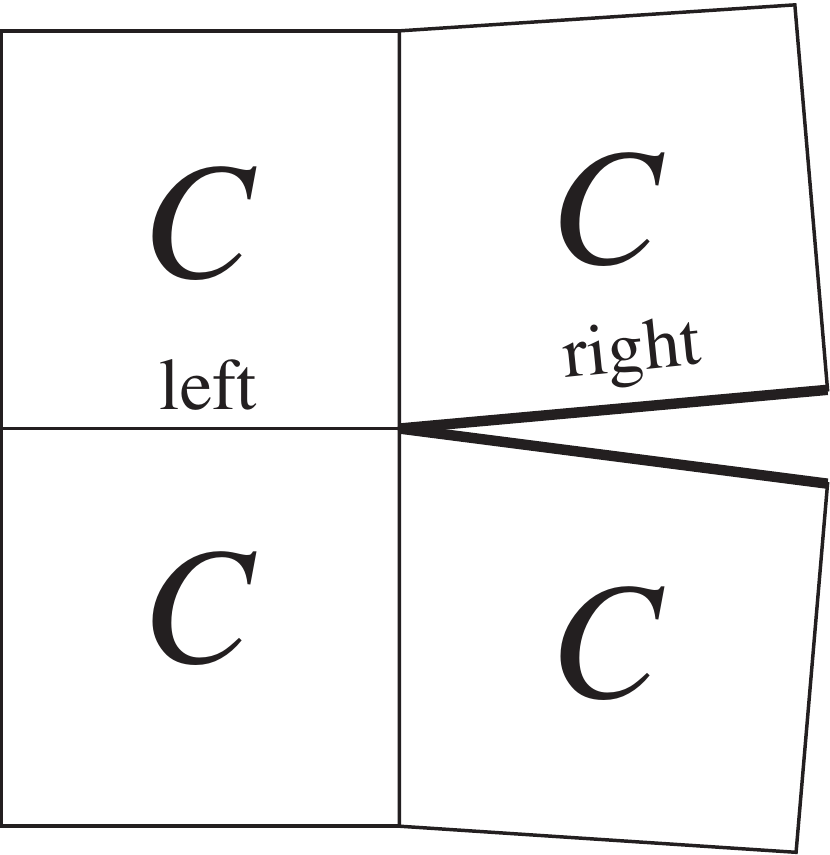}
\caption{The density matrix $\rho$ is represented as the fourth power of CTM $C$.
The configuration sum for the Boltzmann weight of the shown system is taken over all site 
variables, except those on the {\it cut} denoted as `right' on the thicker lines. For this
reason we put index `Right' in $\rho_{\rm Right}^{~}$ when the distinction of left and
	right is necessary.
}
\label{rho}
\end{figure}

Let us consider the six-state clock model on the finite-size square lattice of odd linear dimensions
$L = 2n + 1$, which refer to the number of site variables $\theta_{i,j}^{~}$ positioned at $i^{\rm th}$
row and $j^{\rm th}$ column of the system. The integer $n=1,2,3,\dots$ is associated with CTMRG steps
with which the square lattice gradually expands.
We assume either the free boundary conditions, where all the boundary sites can take
arbitrary directions 
$\theta_{i,j}^{~} = \left\{0, \frac{\pi}{3}, \dots, \frac{5\pi}{3} \right\}$, 
or the fixed ones, where all the boundary sites are fixed to $\theta = 0$.

The system can be divided into four pieces, the quadrants, which are called ``corners'', whose linear 
dimension is $n + 1$. Taking the configuration sum for the Boltzmann weight of each corner and leaving
the site variables being not summed up on the section of the quadrants, one obtains the corner transfer
matrix $C$. The partition function of the system 
\begin{equation}
 Z( L ) \, = \, {\rm Tr} \, C^4_{~} \, = \, {\rm Tr} \, \rho
\end{equation}
is then expressed as the trace of the density matrix $\rho$, which is the fourth power of
$C$. Figure~\ref{rho} shows the corresponding geometry of $\rho$, where the matrix indices 
of $\rho$ correspond to the group of spin variables on upper and lower sides of the cut, shown on 
the right side. By the end of this section, we denote the left and the right side, respectively, by
the indices  `Left' and `Right' when the distinction of the sides is necessary. Thus we express the density matrix, 
shown in Fig.~\ref{rho}, by the notation $\rho_{\rm Right}^{~}$.

Consider $\Psi = C^2_{~}$, which is the Boltzmann weight for the lower (or the upper) half of the system. 
The quantum-classical correspondence suggests that $C^2_{~}$ can be identified as the 
{\it wave function} $\Psi$ of a 1D quantum system of the size $L$; let us denote the corresponding 
quantum state by $| \Psi \rangle$. Since we are considering statistical probabilities, all the 
weights in $| \Psi \rangle$ are real positive numbers. The density operator ${\hat \rho}_{\rm Right}^{~}$, 
whose matrix representation is $\rho_{\rm Right}^{~}$, can be expressed as
\begin{equation}
{\hat \rho}_{\rm Right}^{~} \, = \, {\rm Tr}_{\rm Left}^{~} \, | \Psi \rangle \langle \Psi | \, ,
\end{equation}
where ${\rm Tr}_{\rm Left}^{~}$ denotes the partial trace for the left half of the effective 
1D quantum system. Thus, we can naturally extend the definition of the entanglement 
entropy
\begin{eqnarray}
S( L, T )  
&=&  
 - \, {\rm Tr} \, \left( {\hat \rho}_{\rm Right}^{~} \, \ln \,  {\hat \rho}_{\rm Right}^{~} \right)  
 \nonumber\\
&=& 
- \, {\rm Tr} \, \left(  \rho_{\rm Right}^{~} \, \ln \, \rho_{\rm Right}^{~} \right)
\end{eqnarray} 
to the six-state clock model and also to a wide class of 2D statistical models. 

In the numerical calculation by means of CTMRG, tiny eigenvalues of $\rho$ are discarded 
in the process of renormalization group transformation, and only $\chi$ numbers of block spin states 
are kept. As a result, the calculated entanglement entropy is slightly smaller than $S( L, T )$ defined
in Eq.~(4). In our numerical analysis, we precisely obtain the entanglement 
entropy $S( L, T )$ for relatively small system size up to $L \le 129$, by keeping sufficiently large 
number of block-spin states, $\chi = 300$ at most; under the condition that the numerical cut-off effect 
on $S( L, T )$ is negligible.

When the correlation length $\xi$ of the system is much shorter than $L$, $S( L, T )$ is nearly 
proportional to $\ln \, \xi$ and independent on $L$. On the other hand, when $L \ll \xi$ ,
the entropy $S( L, T )$ is nearly proportional to $\ln \, L$. Thus, at certain temperature $T$,
where the system is critical in the thermodynamic limit, $S( L, T )$ shows a logarithmic divergence
with respect to $L$.

\section{Numerical Results}

\begin{figure}
\includegraphics[width=0.45\textwidth,clip]{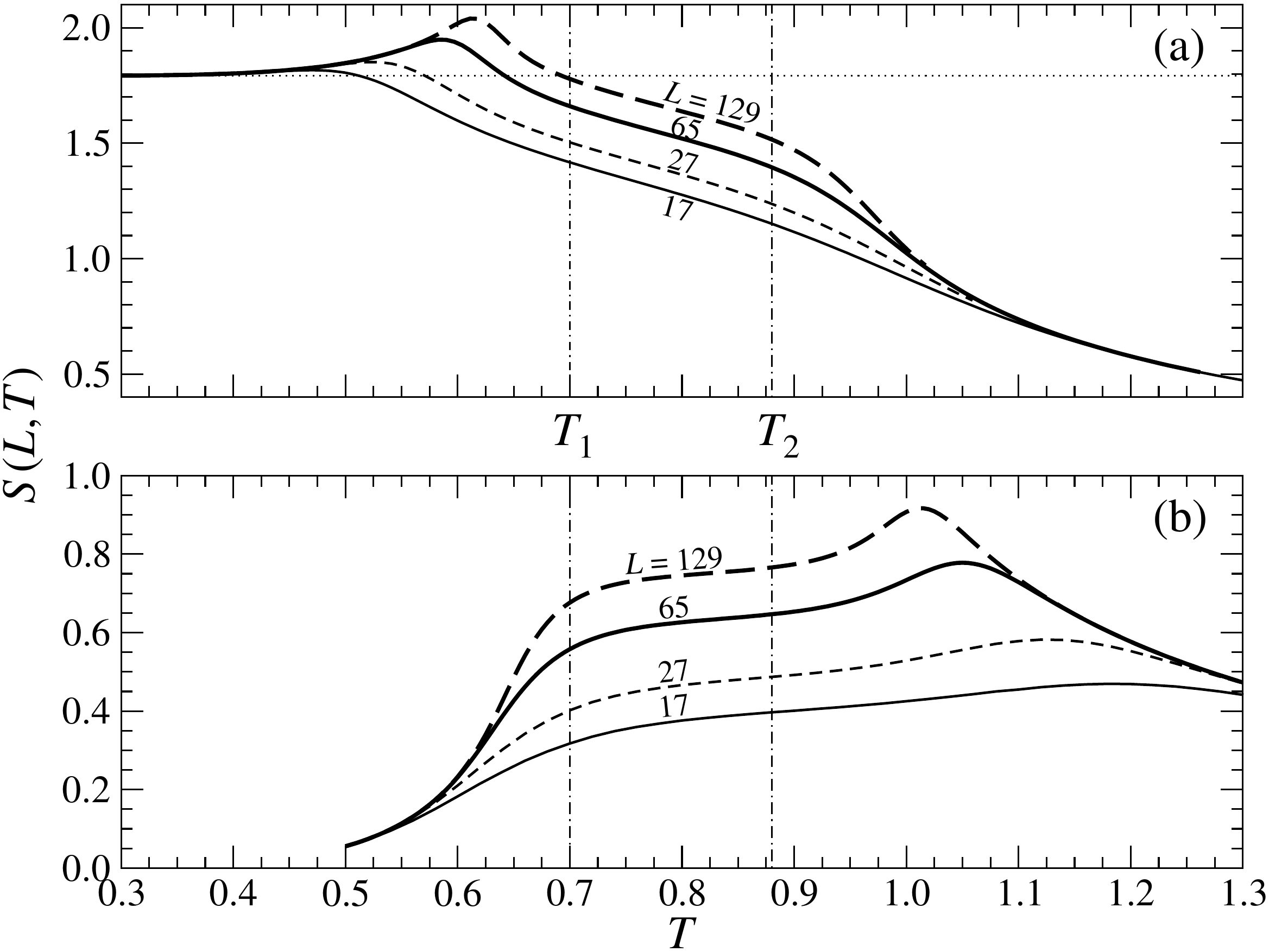}
\caption{Calculated entanglement entropy $S( L, T )$:
(a) under the free boundary conditions, and (b) under the fixed boundary conditions. 
The horizontal dotted line in (a) is the asymptotic value of $S( L, 0 ) = \ln \, 6$.
The two vertical dot-dashed lines at $T_1^{~} = 0.70$ and $T_2^{~} = 0.88$ show the estimated 
transition temperatures by means of finite-size scaling. Sizes of the system are 
specified by the type of the lines; $L=17$ (full thin line),  $L=27$ (dashed thin line), 
$L=65$ (full thick line), and $L=129$ (dashed thick line).}
\label{entropy}
\end{figure}

The temperature dependence of the $S( L, T )$ is shown in Fig.~\ref{entropy} for $L=17,27,65$ 
and $129$. Under the free boundary conditions, as shown in the upper panel (a), 
$S( L, T )$ is equal to $\ln \, 6$ at $T = 0$, and has a peak at $T_1^{*}( L )$, which is an 
increasing function of $L$. Above $T_1^{*}( L )$ the entropy decreases with
$T$, where there is a shoulder on high temperature side. 
Under the fixed boundary conditions, as shown in the lower panel (b), $S( L, 0 )$ is zero.
There is another shoulder in the low temperature side, and a peak at $T_2^{*}( L )$, which is 
a decreasing function of $L$. 

The boundary-condition dependence in $S( L, T = 0 )$ can be explained if we consider the 
corresponding state $| \Psi \rangle$ in Eq.~(3). Under the free boundary conditions,
$| \Psi \rangle$ is a kind of GHZ state~\cite{ghz}, where the density operator
${\hat \rho}_{\rm Right}^{~}$ 
represents a mixed state with 6 equally weighted cases. On the other hand, under
the fixed boundary conditions, $| \Psi \rangle$ is a completely polarized state, and 
the corresponding ${\hat \rho}_{\rm Right}^{~}$ represents a pure state.

\begin{figure}
\includegraphics[width=0.45\textwidth,clip]{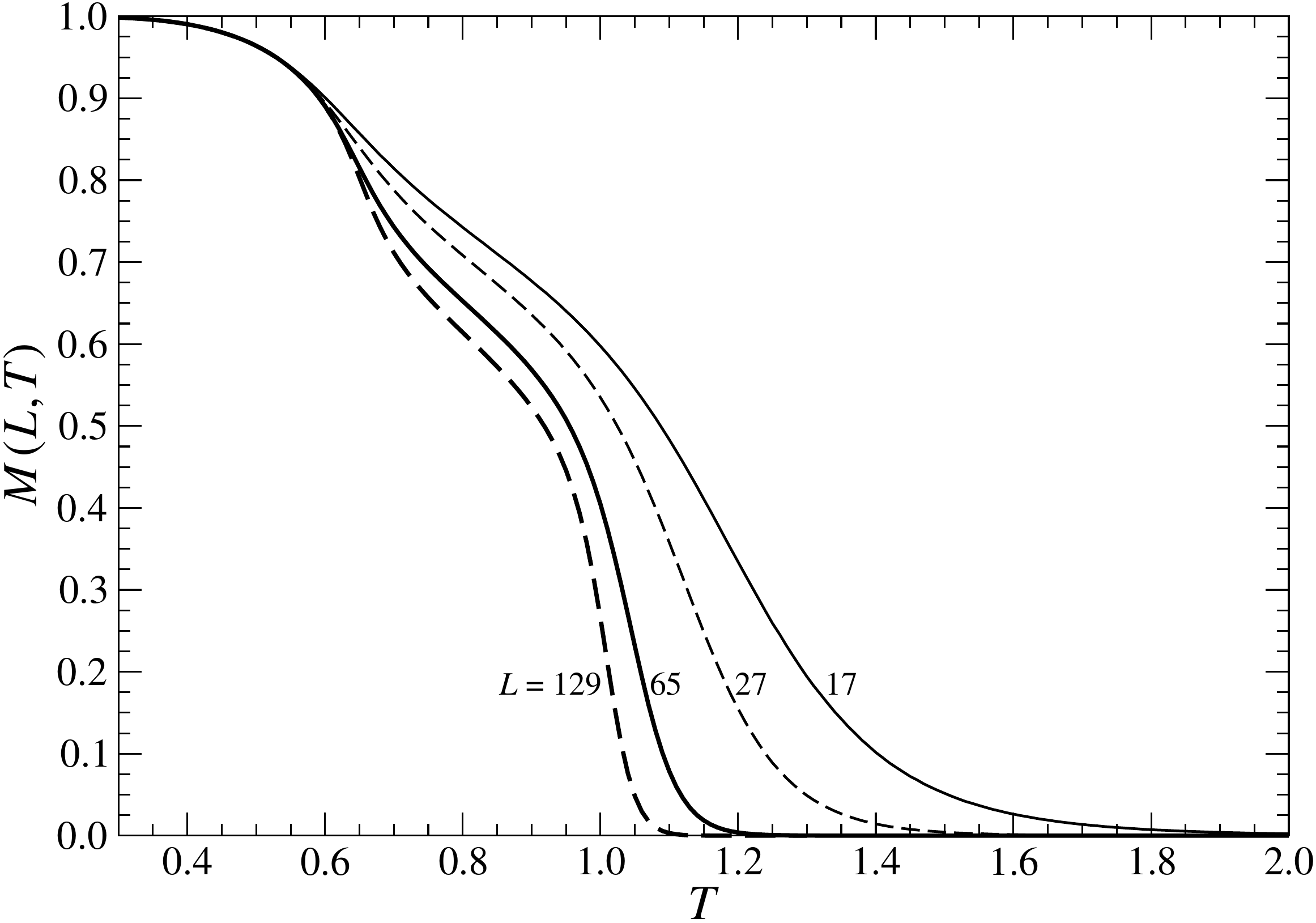}
	\caption{Magnetization $M( L, T ) = \langle \cos ( \theta_{\rm c
}) \rangle$ 
at the center of the system for $L = 17, 27, 65$ and $129$
as a function of temperature calculated under the fixed boundary conditions.
The types of the lines are identical to those in Fig.~\ref{entropy}.}
\label{fig:mag}
\end{figure}

Figure \ref{fig:mag} shows the temperature dependence of the magnetization
\begin{equation}
M( L, T ) = \langle \cos (\theta_{\rm c
})\rangle={\rm Tr}\,\left[\cos (\theta_{\rm c
}) \, 
\rho_{\rm Right}^{~}\right]
\end{equation}	
at the center of the system, where the suffix ${\rm c}$ denotes  
that both $i$ and $j$ are $\frac{L+1}{2}$, under the fixed boundary conditions. 
We have chosen the lattice sizes $L = 17, 27, 65$, and $129$.
We found a good correspondence between the
shoulders in $M( L, T )$ and the peak position $T_1^{*}( L )$ and $T_2^{*}( L )$.

The BKT transition temperatures $T_1^{~}$ and $T_2^{~}$ 
can be obtained by applying the finite-size scaling for 
$T_1^{*}( L, T )$ and $T_2^{*}( L, T )$, respectively 
toward the thermodynamic limit $L \rightarrow \infty$. Let us check this conjecture.
It has been accepted that the correlation length $\xi$ around a BKT transition 
temperature $T_{\rm C}^{~}$ is asymptotically expressed as
\begin{equation}
\xi \, \propto \, \exp\left[ const. \frac{\sqrt{T_{\rm C}^{~}}}
	{\sqrt{ \left| T - T_{\rm C}^{~} \right|}} \right] \, .
	\label{xsi}
\end{equation}
When the system size $L$ is smaller than the right-hand-side of Eq.~\eqref{xsi}, the correlation length is effectively 
suppressed down to $L$. Under such a geometrical constraint, 
it is possible to introduce an effective temperature $T^{*}_{~}( L )$ that satisfies the relation
\begin{equation}
L \, \propto \, \exp\left[ const. \frac{\sqrt{T_{\rm C}^{~}}}{\sqrt{ \left| T^{*}_{~}( L ) - T_{\rm C}^{~} \right| }} \right] \, .
\end{equation}
Solving this relation with respect to $T^{*}_{~}( L )$, we obtain 
\begin{equation}
\label{eq:templog}
	T^{*}_{~}( L ) = T_{\rm C}^{~} + \frac{\alpha}{{\left[ \ln ( \beta L ) \right]}^2_{~}} \, ,
\end{equation}
where $\alpha$ and $\beta$ are appropriate constants. Since the entanglement entropy is 
almost proportional to the logarithm of the correlation length, we can expect an analogous
discussion for the entanglement entropy, too. Let us check whether $T_1^{*}( L )$ and
$T_2^{*}( L )$ satisfy the scaling form in Eq.~(8) or not.

\begin{figure}
\includegraphics[width=0.45\textwidth,clip]{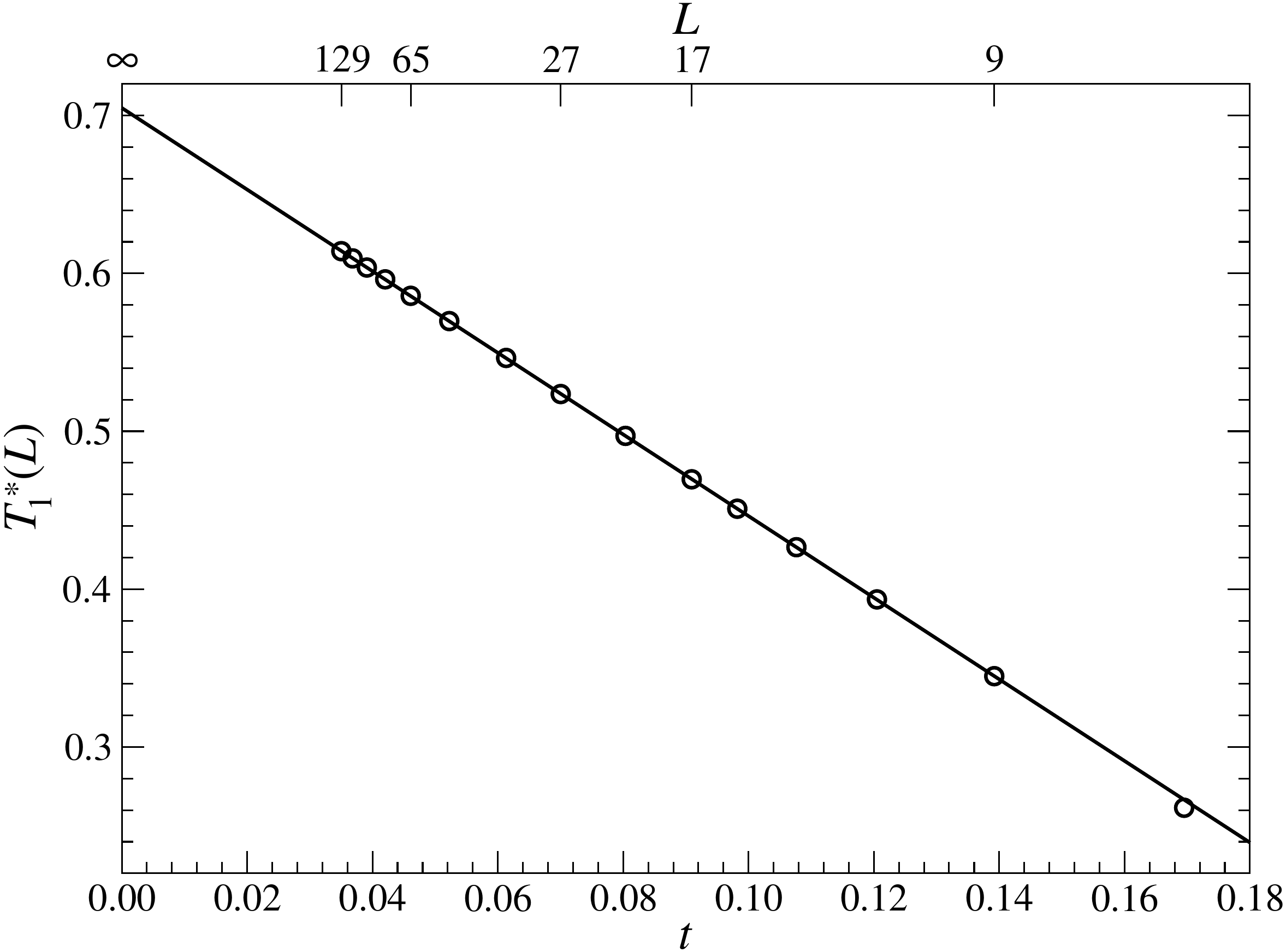}
\caption{Finite-size scaling for the peak position $T_1^{*}( L )$ with respect to $t$ in 
Eq.~(\ref{eq:templog}), which draws $T_1^{*}( \infty ) = 0.70$.}
\label{T1}
\end{figure}
\begin{figure}
\includegraphics[width=0.45\textwidth,clip]{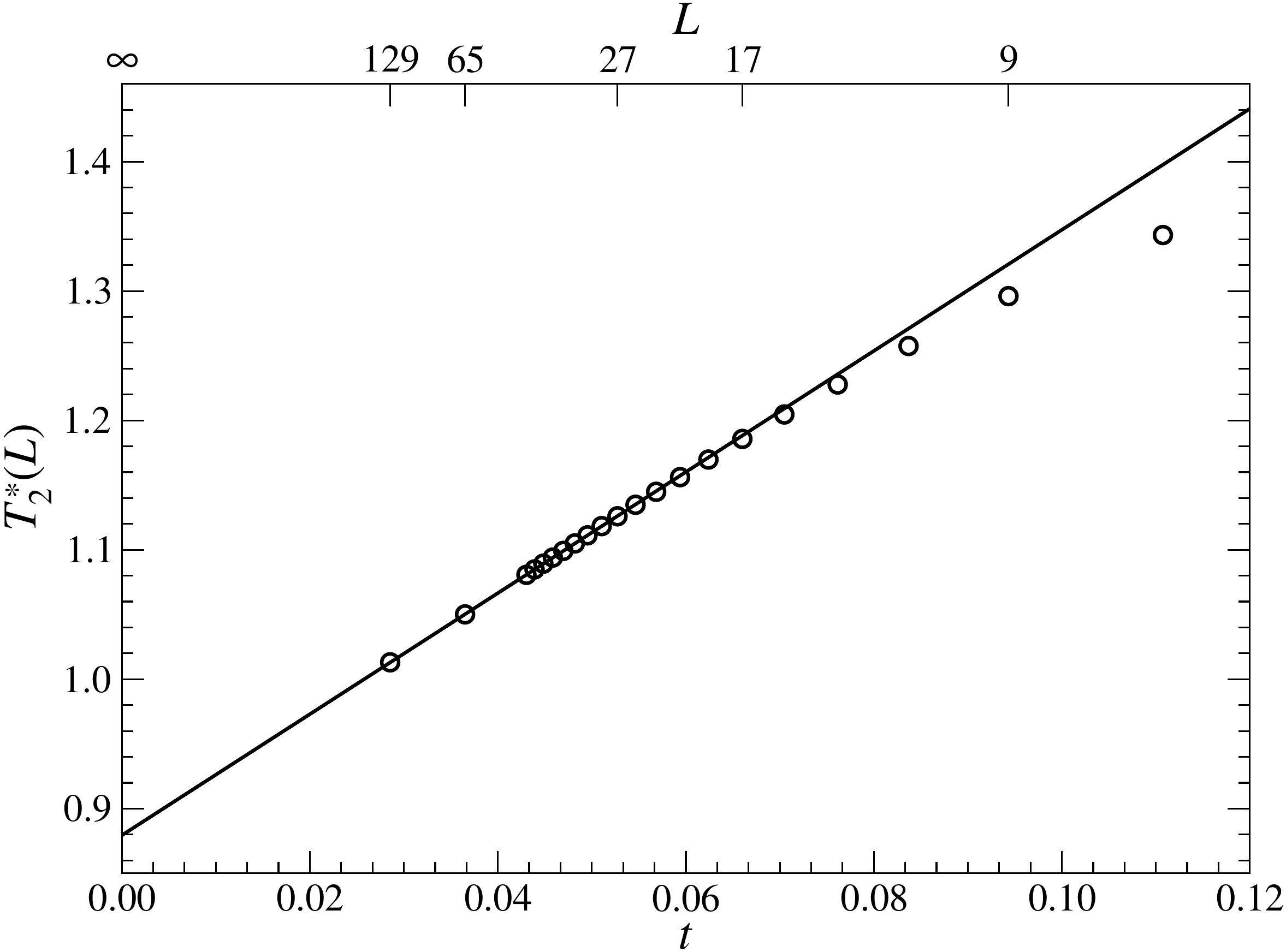}
\caption{Finite-size scaling for the peak position $T_2^{*}( L )$ with respect to $t$ in 
Eq.~(\ref{eq:templog}), which draws $T_2^{~} = 0.88$.}
\label{T2}
\end{figure}

Figure~\ref{T1} shows $T_1^{*}( L )$ with respect to
\begin{equation}
t = \frac{1}{{\left[ \ln (\beta L) \right]}^2_{~}} \, ,
	\label{t-scl}
\end{equation}
where the parameter is chosen as $\beta = 1.62$, which is determined
so that the plots in the figure become linear in $t$ when it is small.
From the slope we obtain $\alpha = -2.58$, and under these parameterizations
we found $T_1^{*}( \infty ) = 0.70$. 
Figure~\ref{T2} shows $T_2^{*}$ with respect to $t$, where the parameter is chosen
as $\beta = 2.88$. From the slope we obtain $\alpha = 4.68$, and
under these parameterization we have $T_2^{~}( \infty ) = 0.88$.

Let us confirm the finite-size scaling by observing the temperature dependence 
of $S( L, T )$ around $T_2^{*}( L )$ directly. Figure~6 shows $S( L, T ) / \ln L$
with respect to the rescaled temperature
$
\left[ T -  T_2^{*}( L ) \right] {\left[ \ln (\,\beta\, L) \right]}^2_{~}
$.
As we can see, the curves slowly converge to a single peak at $ T_2^{*}( L )$.

\begin{figure}
\includegraphics[width=0.45\textwidth,clip]{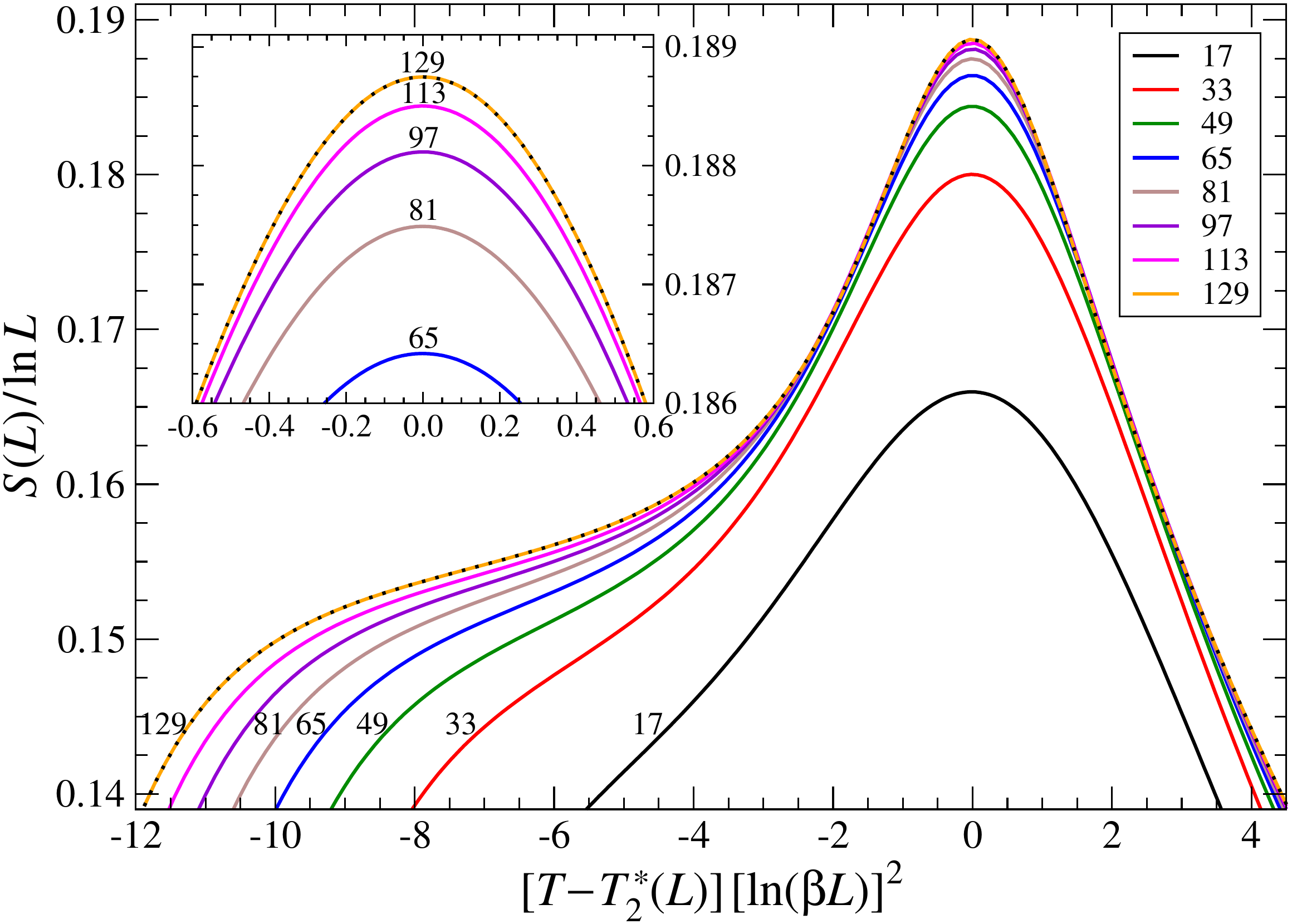}
\caption{(Color online) Scaling plot for $S( L, T )$ according to the rescaled 
	temperature in Eq.~\eqref{t-scl}.}
\end{figure}

\section{Conclusions and Discussions}

We have studied the six-state clock model by means of the CTMRG method and observed the
temperature dependence of the entanglement entropy $S( L, T )$, where $L$ is its system
size. Under the free boundary conditions, the calculated entropy exhibits a peak 
at $T_1^{*}( L )$, which is an increasing function of $L$. Under the fixed 
boundary conditions, the peak is observed at $T_2^{*}( L )$, which is the decreasing
function of $L$. The finite-size scaling on $T_1^{*}( L )$ and $T_2^{*}( L )$, according to the
BKT form of the finite-size correction, draws the final results $T_1^{*}( \infty ) = 0.70$ and 
$T_2^{*}( \infty ) = 0.88$, which agree with the transition temperature reported so far,
as listed in Table~I. 

In our previous study of the $q$-state clock model on hyperbolic lattices, we observed 
a second order phase transition except for $q = 3$, where the location of the transition
point is close to $T_2^{~}$ for each value of $q$, 
and a cross-over behavior around $T_1^{~}$~\cite{gendiar}. On the hyperbolic
lattice, there is no BKT transition, since the hyperbolic nature of the lattice geometry
prevents the divergence of the correlation length. A question arises when we put the
six-state clock model on a weakly curved hyperbolic lattice~\cite{gendiar2}. It would be
possible to estimate both $T_1^{~}$ and $T_2^{~}$ according to a kind of ``finite-curvature scaling''. 

On the nature of the phase transition in the five-state clock model, there is an argument
from the view point of the stiffness~\cite{baek,lapilli,chatelain}: 
If the model exhibits the BKT transition, the width of the critical temperature region is
very small. We conjecture that observation from the entanglement entropy would
be of use also in this case.

\begin{acknowledgments}
	This work was supported by the projects APVV-14-0878 (QETWORK) and VEGA-2/0130/15.
T.~N. and A.~G. acknowledge the support of Grant-in-Aid for Scientific Research.
R.~K. acknowledge the support of Japan Society for Promotion of Science P12815.
\end{acknowledgments}

\end{document}